\documentclass[11pt]{article}

\usepackage[T1]{fontenc}
\usepackage[utf8]{inputenc}
\usepackage{lmodern}
\usepackage[margin=1in]{geometry}
\usepackage{amsmath,amssymb}
\usepackage{bm}                 
\usepackage{textcomp}
\usepackage{gensymb}            
\usepackage{graphicx}
\usepackage{float}
\usepackage[labelfont=bf,font=small]{caption}
\usepackage{setspace}
\usepackage[super,comma,sort&compress]{natbib}  

\usepackage[colorlinks=true,
            linkcolor=blue,
            citecolor=blue,
            urlcolor=blue,
            filecolor=blue]{hyperref}

\newcommand{\Bapp}{B_{\mathrm{app}}}
\newcommand{\Oc}{\Omega_{\mathrm{c}}}
\newcommand{\mzero}{\mu_{0}}
\newcommand{\Am}{\mathrm{A\,m^{2}}}
\newcommand{\gl}{\gamma_{l}}

\begin{document}

\begin{center}
{\LARGE\bfseries Partitioning the magnetization raises the stability limit of ferromagnetic colloids \par}

\vspace{1em}

{\large Paramita Modak\textsuperscript{1},
Priyadharshini~M\textsuperscript{1},
Ankur Jaiswar\textsuperscript{1},
Ambarish Ghosh\textsuperscript{1,2,$\ast$}\par}

\vspace{0.6em}

\begin{minipage}{0.9\linewidth}\centering\small
\textsuperscript{1}Centre for Nano Science and Engineering, Indian
Institute of Science, Bangalore 560012, India\\[2pt]
\textsuperscript{2}Department of Physics, Indian Institute of Science,
Bangalore 560012, India\\[2pt]
\textsuperscript{$\ast$}Correspondence:
\href{mailto:ambarish@iisc.ac.in}{ambarish@iisc.ac.in}
\end{minipage}
\end{center}

\vspace{0.8em}

\noindent\textbf{Abstract}

\noindent
Magnetic actuation scales with a particle's magnetic moment, yet practical colloidal suspensions remain superparamagnetic or weakly magnetized, a permanent dipole drives a near contact attraction growing as the square of the moment, binding particles irreversibly. Here we show that this constraint depends not only on how much magnetic material a particle carries, but on where it sits. Partitioning a fixed moment M into N internal elements leaves the far field interaction, and every actuation force, unchanged while suppressing the near-field pair energy by 1/N. As we show in our experiments across 57 samples and three geometries, stability collapses onto a single $\sim M^{2}/N$ boundary, raising the accessible dipolar energy scale nearly 150-fold. A liquid-phase analogue of magnetic force microscopy images the engineered energy landscape directly. The resulting colloids outrun size-matched superparamagnetic beads in weak fields far from a magnetic source, suggesting their applicability in hard to reach places. Together, these results open a new operating regime for colloidal manipulation, with impactful implications in microfluidics and drug delivery.

\section*{Introduction}
\label{sec:intro}

Among ways to exert forces on colloidal matter deep
inside optically opaque, biologically crowded environments, magnetic
fields are unmatched in penetration and biocompatibility, which places
magnetic colloids at the heart of technologies spanning
microfluidics\cite{ref1,ref2}, sorting\cite{ref3}, drug
delivery\cite{ref4,ref5}, hyperthermia and imaging\cite{ref6,ref7},
magnetically guided implants\cite{ref8}, and micro and
nanorobotics\cite{ref9,ref10,ref11,ref12,refAG1,refAG2,ref13}. Beyond these specialized settings, magnetic bead manipulation has become a fixture of routine laboratory practice. The force
exerted on a particle of moment $\bm{m}$ in a field gradient is
$\bm{F}=(\bm{m}\cdot\nabla)\bm{B}$, and the torque in a rotating field is
$\bm{\tau}=\bm{m}\times\bm{B}$, with associated energy scale $mB$. Every
actuation task therefore improves with moment. Yet the
agglomeration-free colloids in practical use are superparamagnetic
(SPM), whose induced moment collapses as the field
weakens\cite{ref23,ref24,ref27,ref37}, or ferromagnetic with
deliberately small permanent moments. The cost is felt precisely where
magnetic manipulation is most valuable: nanomotors with limited thrust,
gradient forces that fade far from the field source, and on-demand
control that cannot be triggered in deep, weak-field regions.

The obstacle to simply raising the moment was articulated by de Gennes
and Pincus~\cite{ref51}: two FM colloids of size $d$ and moment $m$
attract head-to-tail with dipolar energy
$U_{\mathrm{int}}\propto\mzero m^{2}/2\pi d^{3}$. This attraction grows
quadratically with the moment, so in the high-moment regime that
actuation favours, it overwhelms the electrostatic barrier that
stabilizes ordinary colloids at near contact, thus driving them irreversibly
into chains and clusters~\cite{chain1,chain2}. A bound pair can in principle be pulled
apart by the torque of a time-varying field \cite{deagglomeration}, but the required amplitude
grows with the binding energy and becomes impractically large for
high-moment aggregates. Half a century of magnetic colloid design has
consequently treated the near-contact attraction as fixed by $m$ and
$d$, and worked around it: reducing the moment, switching to induced
(SPM) moments, or padding particles with non-magnetic
coatings\cite{spacer}. High-moment ferromagnetic colloids have
consequently remained inaccessible as stable suspensions.

Here we show that the near-contact attraction is not fixed by $m$ and
$d$ alone. It depends, in addition, on where the magnetization resides inside
the particle, and that internal degree of freedom can be engineered
independently of the total moment. Partitioning a fixed moment $M$
across $N$ thin ferromagnetic elements leaves the far-field interaction
unchanged, because at separations beyond the particle size the elements
act collectively as a single dipole $M$, while suppressing the
near-field pair energy by $1/N$, because at close approach each element
interacts predominantly with its single facing counterpart. The
de-agglomeration condition $M\Bapp\gtrsim U$ therefore relaxes linearly
with $N$: larger $N$ lowers the field needed to release a bound cluster
while leaving the actuation force on a free colloid untouched.
Experimentally, this yields stable FM colloids whose accessible dipolar
energy scale is nearly 150-fold beyond that of an unpartitioned
reference.

Distributing functional, including magnetic material within a particle is itself familiar:
multilayered nanowires exploit it for magnetoresistive
transport\cite{physical2}, segmented rods for optical
barcoding\cite{physical3}, and Fe--silica composites for
superparamagnetic response\cite{refPhilipse,refTartaj}. In every case,
however, the internal distribution is used to set a single-particle
property. Crucially, this internal architecture also dictates inter-particle interactions: the
magnetization landscape decouples the far-field moment, which sets
actuation, from the near-field attraction, which sets stability. This
decoupling separates the present approach from both established
stabilization routes, each of which pays a price the application feels.
Superparamagnetic composites suppress the attraction by giving up the
permanent moment altogether; spacer layers and thick
coatings\cite{spacer} suppress it only by forcing the dipoles apart, so
that matching a $1/N$ suppression inflates the particle volume by a
factor $\sim N^{2/3}$ and with it the hydrodynamic drag that limits
actuation (\hyperref[sec:discussion]{Discussion}).
Unlike these traditional stabilization strategies, internal partitioning preserves the external dimensions and low-drag geometry, on which magnetic propulsion depends, and in the fully dispersed limit
($N\!\to\!\infty$) the interaction additionally softens from $1/d^{3}$
to a bounded $1/d^{2}$.

We establish the principle analytically and numerically, and demonstrate
it experimentally across three anisotropic geometries: helices, bent
rods and straight rods, with layer counts from $N=1$ to $36$ and, via
co-evaporation, in the $N\!\to\!\infty$ limit. A stability phase diagram
spanning 57 samples shows all geometries collapsing onto a single
de-agglomeration boundary that follows the predicted $M^{2}/N$ scaling,
confirming that the controlling variable is the partitioning of the
moment rather than the shape of the particle or the route by which it
was made. Single particle magnetometry based on the step-out frequency
verifies that every architecture retains a permanent moment, ruling out
a superparamagnetic crossover in the thin layers. A liquid-phase
analogue of magnetic force microscopy, using diffusing SPM beads as
probes, maps the force landscape around individual particles and
directly visualizes its engineered flattening. Finally, we demonstrate
these colloids operating in a regime that has remained out of reach:
dispersed, individually addressable, high-moment FM particles that
outrun SPM beads in the weak-field region far from a field source (permanent magnet). To
our knowledge, this is the first demonstration that the internal spatial
distribution of the magnetization serves as a design variable for the
near-field pair potential of magnetic colloids. The 
fifty-year-old constraint is therefore re-framed by accounting for the precise internal localization of the magnetization.

\section*{Results}
\label{sec:results}

\subsection*{Engineering the pair potential through the magnetization
landscape}

\begin{figure}[H]
\centering
\includegraphics[width=\linewidth]{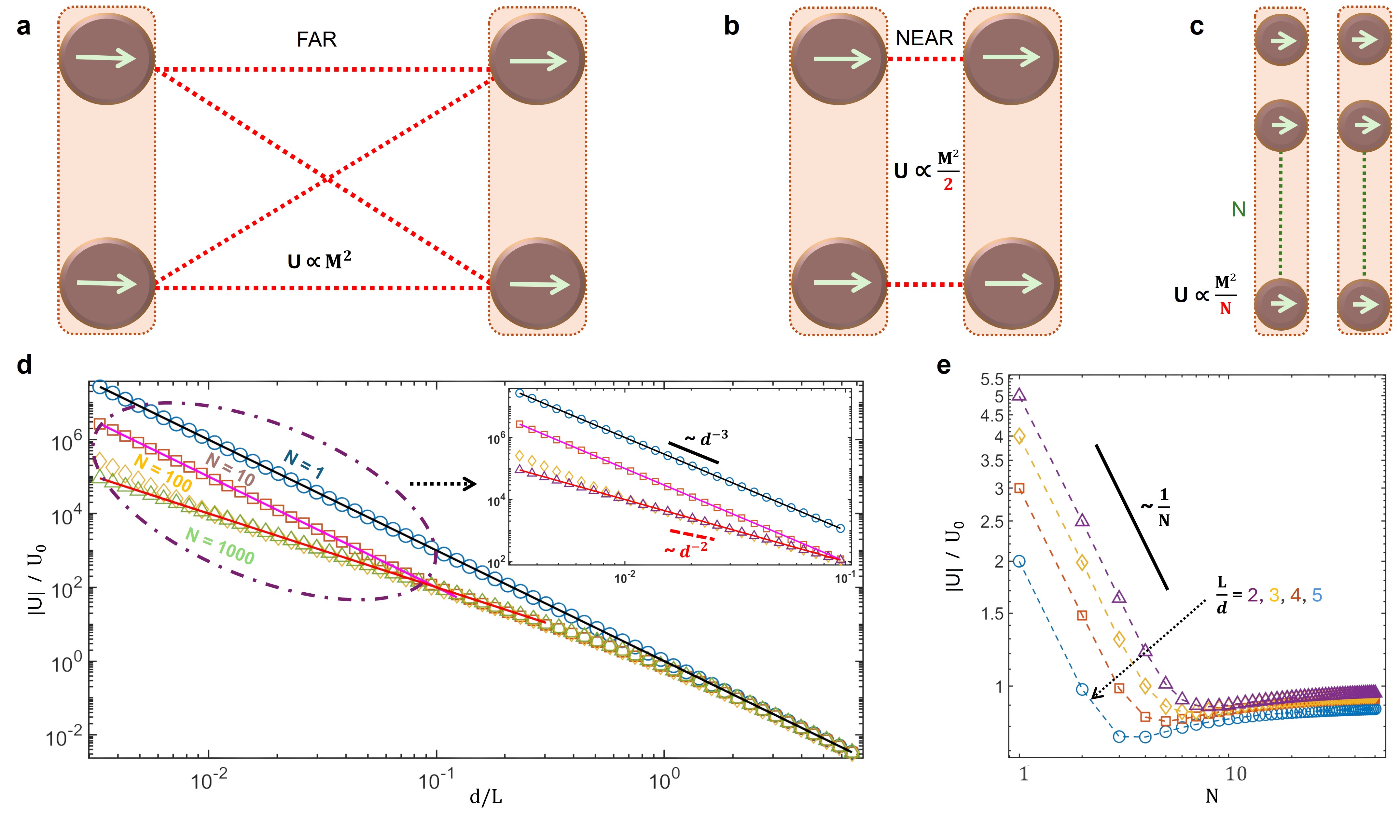}
\caption{\textbf{--- Engineering the pair potential through the internal
magnetization landscape.} \textbf{a}, Far field ($d\gtrsim L$): every
ferromagnetic element of one particle interacts with every element of
its neighbour (red dashed lines); the contributions sum to the
point-dipole result $U\propto M^{2}$, independent of $N$ (shown for
$N=2$). \textbf{b}, Near field ($d\lesssim\Delta l$, separations below
the inter-element spacing): each element interacts predominantly with
its facing counterpart, diagonal contributions being negligible; for
$N=2$ the energy is halved, $U\propto M^{2}/2$. \textbf{c}, For general
$N$ the same counting gives $U\propto M^{2}/N$. \textbf{d}, Computed pair
energy versus separation. In the near-field window ($d/L\lesssim0.1$,
circled), $N=1$ follows $U\propto d^{-3}$ (black) and $N=10$ the same
power law suppressed by $1/N$ (magenta), reflecting discrete pairwise
interactions. At large layer counts ($N=100,1000$) the finely discretized chain behaves as a
continuous line source and the scaling softens to $U\propto d^{-2}$
(red). Inset: magnified near-field region ($d/L\le0.1$) showing the with power law variation clearly. 
The interaction energy is scaled by
$U_0 = {\mu_0 M^2}/{2\pi L^3}$,
plotted against interparticle separation, scaled by the particle length L. \textbf{e}, Scaled interaction energy plotted against the number of magnetic layers $N$, for four fixed chain-length-to-separation ratios $L/d = 2$--$5$.}
\label{fig:1}
\end{figure}

Consider two identical FM colloids, each carrying a total moment $M$
partitioned into $N$ collinear elements of moment $m=M/N$, spaced
$\Delta l$ apart along the particle axis so that the magnetized region
has length $L=N\Delta l$ (Fig.~\ref{fig:1}). Two regimes follow from
comparing the inter-particle separation $d$ with these internal length
scales. In the far field, $d\gg L$, every element of one particle
interacts with every element of the other; the $N^{2}$ pairwise
contributions add coherently and the particles interact as two-point
dipoles of moment $M$, with $U\propto M^{2}/d^{3}$ independent of $N$
(Fig.~\ref{fig:1}a). All actuation forces and torques act in this regime
and are therefore untouched by the partitioning. In the near field,
$d\ll\Delta l$, separations below the spacing between elements,
each element interacts appreciably only with its directly facing
counterpart on the neighboring particle, every other pair being at
least $\Delta l$ away and cut off by the rapid $1/d^{3}$ decay. The
energy is then $N$ pairs, each of strength $(M/N)^{2}$, giving
$U\propto M^{2}/(N d^{3})$, the far-field prefactor divided by $N$
(Fig.~\ref{fig:1}b, c). The suppression is a genuine reduction of the
near-contact energy at fixed external dimensions, not a consequence of increased separation. At intermediate separations, $\Delta l\ll d$, the discreteness is no longer resolved, each particle acts as a
uniformly magnetized thin wire and the interaction takes the bounded,
$N$-independent form $U\propto M^{2}/(L d^{2})$, the signature of
dimensional reduction, a particle with fully dispersed magnetization
interacting as a line dipole rather than a point dipole, matching the
discrete result continuously at $d\approx\Delta l$. At a fixed
near-contact separation $d$, the pair energy therefore falls as $1/N$
until $N\approx L/d$ and then saturates at the thin-wire value: the
$1/N$ suppression terminates in a stable optimum, and the distance
dependence softens from $1/d^{3}$ to $1/d^{2}$ (Fig.~\ref{fig:1}d, e).

The magnetic term enters the total pair potential alongside
contributions that partitioning leaves untouched: for particles of the same size and surface chemistry, the electrostatic (DLVO) repulsion is identical for $N=1$ and $N>1$. Partitioning does not keep particles apart, the far-field attraction is unchanged, so they still approach and fall into contact; what changes is the depth of the well they land in. For an unpartitioned particle that well is deep and the bound state irreversible, whereas partitioning makes it shallower by $1/N$, shallow enough that a modest field can lift the particles back out. The agglomeration therefore becomes reversible, achieving on demand what a superparamagnetic colloid gets by switching the field off, except that the moment here is permanent and release is driven by an oscillating or rotating field rather than by its removal~\cite{deagglomeration}. Release is set not by thermal energy but by the torque energy $M\Bapp$ an applied time-varying field delivers per particle: de-agglomeration requires $M\Bapp\gtrsim U$. Because $U$ falls as $1/N$ at fixed $M$, so does the required field; equivalently, for a given maximum achievable field the usable moment rises as $M_{\max}=(2\pi d^{3}/\mu_0)\,N\,B_{\mathrm{app}}^{\max}$ while the layers
remain resolved ($N\lesssim L/d$), then saturates at the bounded
thin-wire value $(2\pi d^{2}L/\mu_0)\,B_{\mathrm{app}}^{\max}$.
Partitioning thus raises the ceiling on stable moment up to this
optimum, while the actuation force on the freed particle still scales
with the full $M$. While real ferromagnetic layers have finite thickness, possible domain structure and shape-dependent demagnetizing fields; our simple point-dipole treatment manages to provide reasonable quantitative agreement with the experiments below without micromagnetic simulation.

\subsection*{Three geometries, one architecture variable}

To test the prediction that partitioning the moment converts
irreversible agglomeration into reversible, we fabricated three
anisotropic geometries by glancing-angle deposition
(GLAD)\cite{ref25,refGLAD} onto seeded substrates: (i) helical
microstructures ($\sim3$~\textmu m helical length, $\sim600$~nm outer
diameter); (ii) bent rods ($\sim3$~\textmu m length, $\sim600$~nm
diameter, with a controlled bend that breaks rotational symmetry and
enables rotating-field propulsion\cite{bentrod}); and (iii) straight
rods ($\sim1$~\textmu m length, comparable diameter), the most direct
analogue of conventional magnetic colloids.

\begin{figure}[H]
\centering
\includegraphics[width=\linewidth]{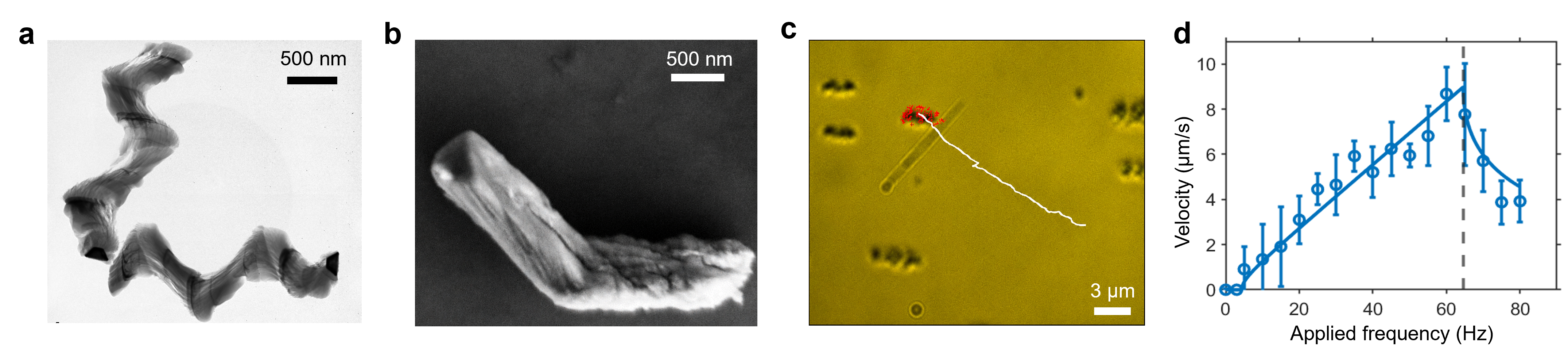}
\caption{\textbf{--- Anisotropic ferromagnetic geometries and
per-particle magnetometry.} \textbf{a}, Representative SEM of helical
microstructures, showing the Fe/SiO\textsubscript{2} layer architecture.
\textbf{b}, Bent-rod SEM image showing the same multilayer construction.
\textbf{c}, Trajectory of a helical microstructure under an applied
rotating magnetic field (10~G, 30~Hz). \textbf{d}, Representative
measurement: velocity versus rotating-field frequency for a helix; the
dashed line marks the step-out frequency from which the magnetic moment
is extracted.}
\label{fig:2}
\end{figure}

In each geometry the magnetic
component is deposited either as a single thick Fe layer ($N=1$), as a
stack of alternating Fe/SiO\textsubscript{2} layers (multilayer
architecture, $N=2$--$36$), or as a co-evaporated Fe/SiO\textsubscript{2}
composite in which Fe nanoislands are dispersed throughout the
SiO\textsubscript{2} matrix, realizing the $N\!\to\!\infty$ limit
(Fig.~\ref{fig:2}a, b).

Quantifying the stability boundary requires the moment of individual
colloids, not of an ensemble. We obtain it from the step-out frequency
$\Oc$ of each swimmer under a rotating field (Fig.~\ref{fig:2}c, d):
beyond $\Oc$ the structure loses synchronization with the field and the
propulsion velocity peaks, with $\Oc\propto MB/\gl$ and the rotational
drag coefficient $\gl$ known for slender chiral
bodies\cite{ref13,ref30,refGhoshDyn}. This in-fluid, single-particle
protocol is the appropriate magnetometry for individual colloids and is,
for GLAD-grown samples, more reliable than ensemble techniques
(SQUID/VSM), which are contaminated by unshadowed Fe deposited on the
substrate around the structures and consequently overestimate the
per-particle moment\cite{ref29}. The same measurement doubles as a
diagnostic of the magnetic state, as shown below.

\subsection*{Distributing the moment converts irreversible into
reversible agglomeration}

\begin{figure}[H]
\centering
\includegraphics[width=\linewidth]{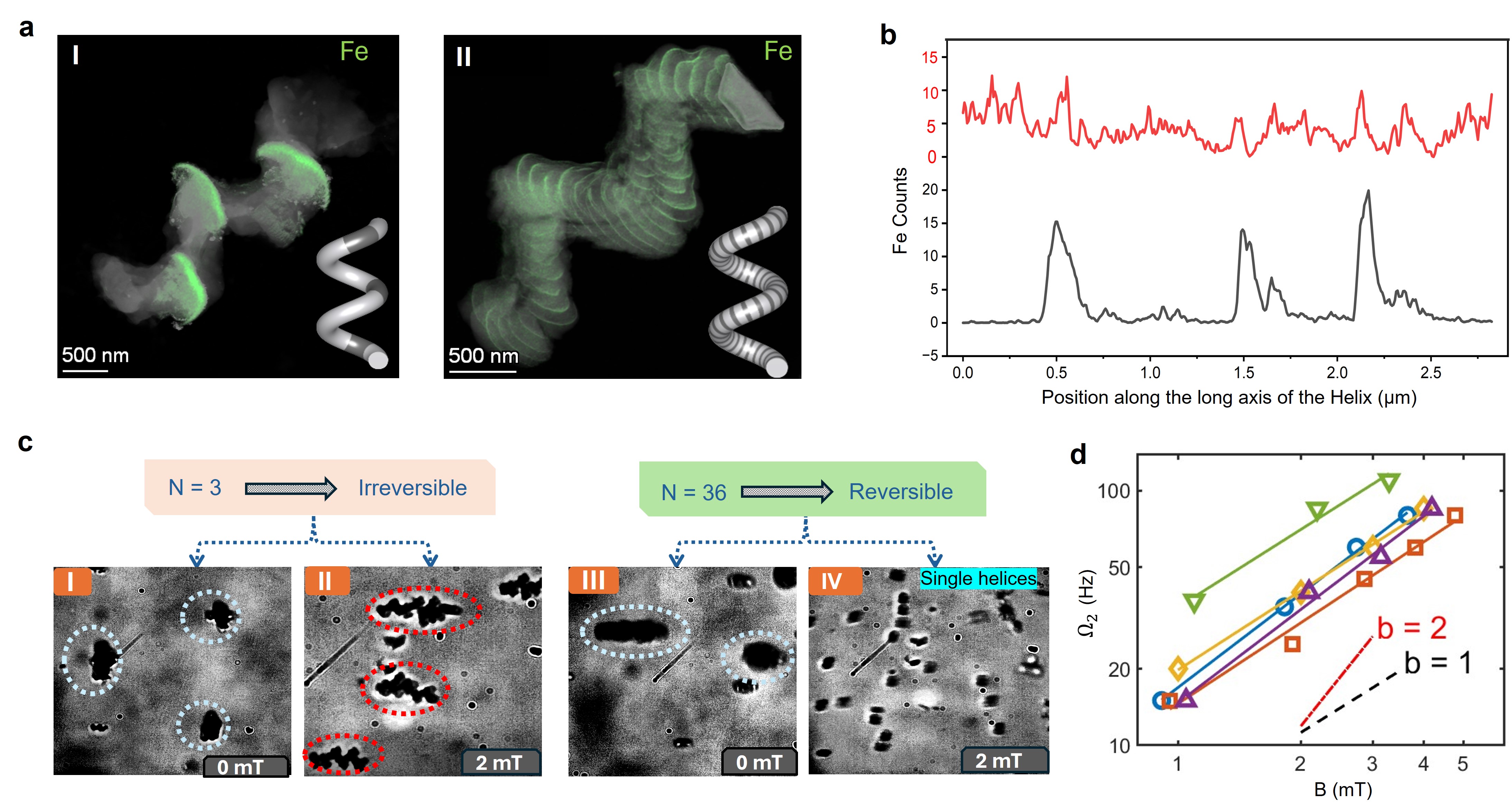}
\caption{\textbf{--- Distributing the moment converts irreversible into
reversible agglomeration; all particles are ferromagnetic.} \textbf{a},
TEM of $N=3$ and $N=36$ helical microstructures with the same total
magnetic moment. \textbf{b}, Elemental Fe line scans (EDS) along the
helical long axis: three localized peaks for the 3-layer structure
($\sim7$~nm Fe each) and high-frequency Fe modulation for the 36-layer
structure. \textbf{c}, Optical images. In zero field both samples form
clusters (I, $N=3$; III, $N=36$). Under a rotating field near step-out
(II, IV), the 3-layer clusters remain locked as rigid rotating
aggregates (Movie~S1), while the 36-layer clusters disassemble into
individually propelling helices (Movie~S2). \textbf{d}, Step-out
frequency $\Oc$ versus applied field $B$ across the full set of
multilayer and co-evaporated samples. The measured scaling
$\Oc\propto B^{1.13\pm0.10}$ is consistent with the slope $b=1$ expected
for a permanent (FM) moment and inconsistent with $b=2$ expected for
induced-moment (SPM) behaviour.}
\label{fig:3}
\end{figure}

Fig.~\ref{fig:3} contains the principal experimental observation of this work. We compare two helical microstructures of nearly identical
geometry and matched measured total moment
($M\approx6\times10^{-15}~\Am$), differing only in how the Fe is
distributed along the helix long axis: three thick layers ($N=3$) versus
a stack of thirty-six thin layers separated by SiO\textsubscript{2}
spacers ($N=36$). Elemental Fe line scans (Fig.~\ref{fig:3}b) confirm
the two architectures: three localized peaks for $N=3$, and a
high-frequency Fe modulation for $N=36$. In zero field both samples form clusters, as thermal motion and the long-range dipolar attraction bring particles into contact, and in both cases the near-contact well exceeds the thermal energy. The diagnostic experiment is what happens under a rotating field near the step-out frequency
($\Bapp\approx10$~G, $65$~Hz). The $N=3$ clusters remain locked as rigid rotating aggregates: the applied torque cannot supply the binding energy (Movie~S1). The $N=36$ clusters disassemble\cite{deagglomeration} at once into individually propelling helices (Movie~S2). Colloids with the
same moment, and therefore the same far-field dipolar energy scale,
exhibit opposite near-field fates; the difference resides entirely in
the internal magnetization landscape.

De-clustering is not specific to rotating fields. Any time-varying field of sufficient amplitude is effective provided the drive frequency $\Omega$ lies between the step-out frequency of the agglomerated cluster, $\Oc^{\mathrm{clust}}$, and that of an isolated microstructure, $\Oc$. Experimentally, we observe $\Oc^{\mathrm{clust}}<\Oc$: the
cluster's rotational drag grows faster with size than its magnetic
moment. Within this window the cluster cannot follow the field while the
individual structures can. Under a rotating field, a phase lag develops
between the field and the internal dipoles, and the resulting internal
torques pull the cluster apart; under an oscillating field, the cluster
cannot reorient between successive half-cycles, and the alternating stress on the dipolar bonds breaks them. The released
microstructures, still below their own $\Oc$, stay locked to the field and propel individually. Clusters whose binding energy exceeds $M\Bapp$ remain agglomerated within the same window.

A natural concern is that thin Fe layers, or the Fe nanoislands of the co-evaporated samples, might cross over to superparamagnetic
behaviour\cite{ref50}, in which case reversibility would be trivial: the moment would simply vanish in zero field. We rule this out directly.
First, every multilayer and co-evaporated sample agglomerates in zero
field, which requires a permanent moment. Second,
across the full sample set the step-out frequency scales with the
applied field as $\Oc\propto B^{1.13\pm0.10}$ (Fig.~\ref{fig:3}d),
consistent with the linear dependence ($b=1$) expected for a permanent
moment and inconsistent with the quadratic dependence ($b=2$) expected
for an induced moment as in SPM-based systems\cite{jyoti,SPM1,SPM2,SPM3,SPM4}. The
reversibility therefore arises from the spatial distribution of the
magnetization, not from any change in the magnetic state of the
material.

\subsection*{A universal stability phase diagram}

\begin{figure}[H]
\centering
\includegraphics[width= 0.80\linewidth]{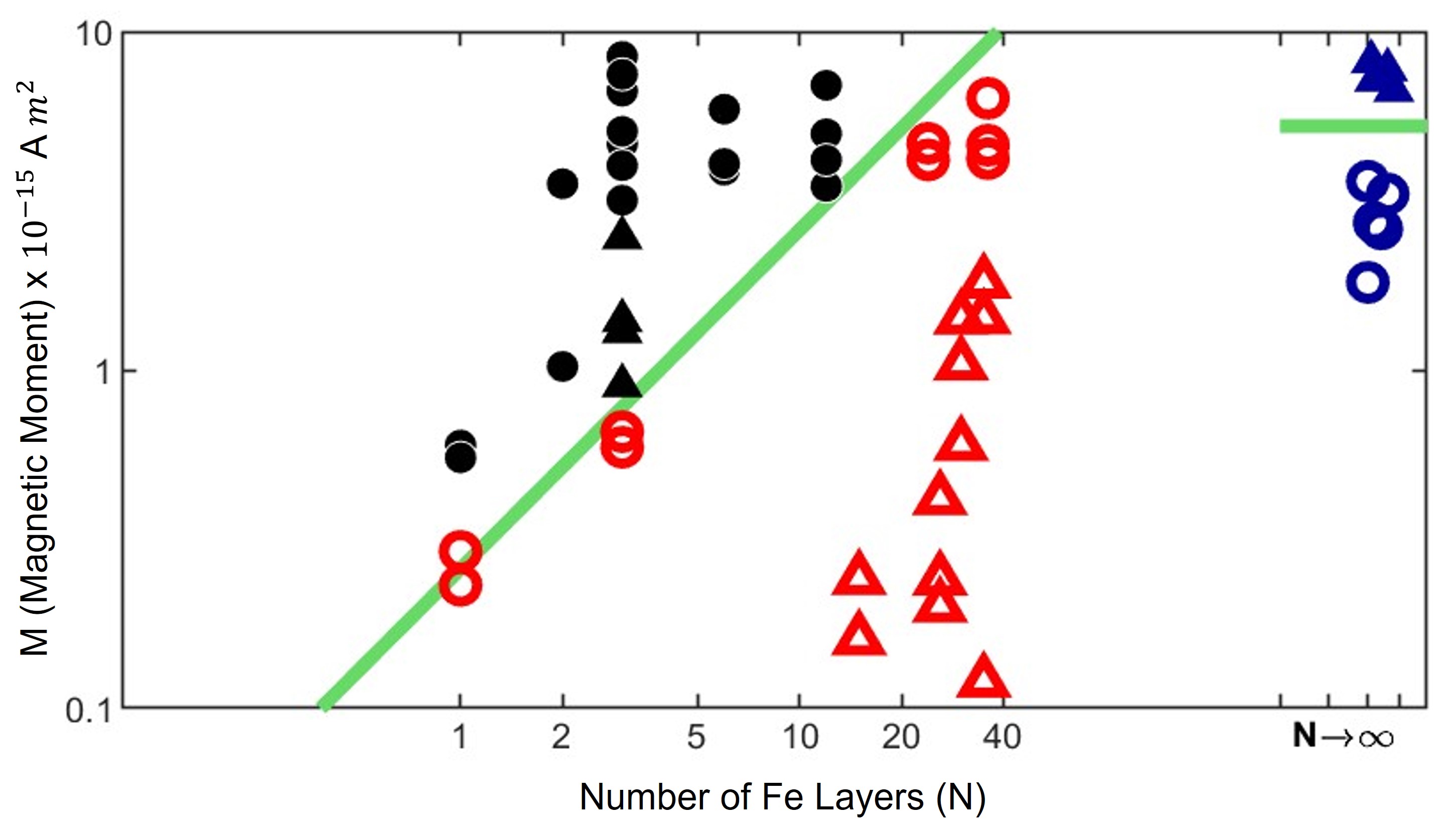}
\caption{\textbf{--- Stability phase diagram.} Layer number $N$ plotted
against the experimentally measured magnetic moment ($\Am$) for 57
helical and bent-rod samples (multilayered and co-evaporated). Open
symbols: reversible agglomeration ($\Bapp\le70$~G). Filled symbols:
irreversible. Circles: helices; triangles: bent rods; blue symbols:
co-evaporated microstructures. Solid green curves: theoretical boundary
from the energy balance $U\approx\mzero M^{2}/2\pi d^{3}N\approx M\Bapp$
with $d=400$--$650$~nm. Co-evaporated samples are plotted near the right
edge of the abscissa at large effective $N$ ($N\!\to\!\infty$), where
the theoretical boundary is $U\approx\mzero M^{2}/2\pi d^{2}L\approx
M\Bapp$.}
\label{fig:4}
\end{figure}

Figure~\ref{fig:4} maps the reversible--irreversible boundary in the
$(N, M)$ plane for 57 samples spanning 13 helical and bent-rod
geometries, layer counts $N=1$--$36$, and co-evaporated samples in the $N\!\to\!\infty$ regime. Open symbols denote reversible agglomeration (clusters disassemble at $\Bapp\le70$~G); filled symbols denote irreversible agglomeration. The theoretical boundary follows from a single energy balance: equating the near-field energy
$U\approx\mzero M^{2}/2\pi d^{3}N$ to the torque energy $M\Bapp$
available per particle, with the contact separation $d$ fixed by the
measured particle diameters ($d=400$--$650$~nm), leaves $\Bapp$ as the only free parameter. The fit returns $\Bapp\sim2$--$10$~G, below the experimentally applied $\lesssim70$~G, as expected: the two-particle model underestimates the binding of many-particle aggregates. 
The conclusion is insensitive to the precise value: as shown earlier,
$M_{\max}=(2\pi d^{3}/\mu_0)\,N\,B_{\mathrm{app}}^{\max}$, so changing $B_{\mathrm{app}}$ merely rescales the boundary toward higher moment without altering the underlying $1/N$ scaling on which the data collapse.

Two features of the diagram carry the central message. First, helices
and bent rods collapse onto a common boundary: the controlling physics is the partitioning of the moment, not the topography of the particle. Second, the diagram quantifies the gain directly: the unpartitioned reference ($N=1$) clusters irreversibly, whereas the $N=36$ architecture remains stable at a moment 12 times larger, a dipolar energy scale ($\propto M^{2}$) nearly 150 times greater. We emphasize that the relevant energy scale for de-agglomeration is the applied-field torque energy $M\Bapp$, not the thermal energy $k_{\mathrm{B}}T$: the diagram shows field-releasable versus locked states. The combination of colloidal stability with enhanced moment bears directly on practical deployment: a high-moment swarm propels faster, exerts larger forces, and penetrates tissue and confined geometries more effectively, expanding the scope of magnetic manipulation in biomedicine and active matter.

The co-evaporated samples extend the same trend to the fully dispersed limit. Their maximum stable moments match those of multilayer samples
with $N\approx30$--$35$, and fitting their boundary with the thin-wire
form $U\approx\mzero M^{2}/2\pi d^{2}L$ returns $\Bapp\sim8$--$26$~G,
consistent with the pair-interaction picture. That the co-evaporated
limit ties, rather than exceeds, the highest multilayer architectures is
itself the predicted behaviour: the $1/N$ suppression saturates as the
near-field energy approaches the bounded thin-wire value, so beyond
$N\sim L/d$ further partitioning brings diminishing returns, the force
landscape is already as flat as the geometry allows. The
three-dimensional dispersal of the Fe islands additionally relaxes the
one-dimensional thin-wire assumption, moderating the suppression
relative to the ideal limit. Monte Carlo
calculations for the realistic geometry, a fixed total moment
distributed over randomly placed Fe islands throughout the cylinder
volume, confirm that the ensemble-averaged pair energy is governed by
the center to center separation and is insensitive to the details of the
random internal arrangement. In summary, more $N$ is
never worse, and co-evaporation offers a lithography free, single-step
route to the optimum.

\subsection*{Direct mapping of the in-situ force landscape}

The stability diagram measures the consequence of the engineered
landscape, which we attempt to image. Magnetic force microscopy has previously been extended from vacuum into liquid using a 

\begin{figure}[H]
\centering
\includegraphics[width=\linewidth]{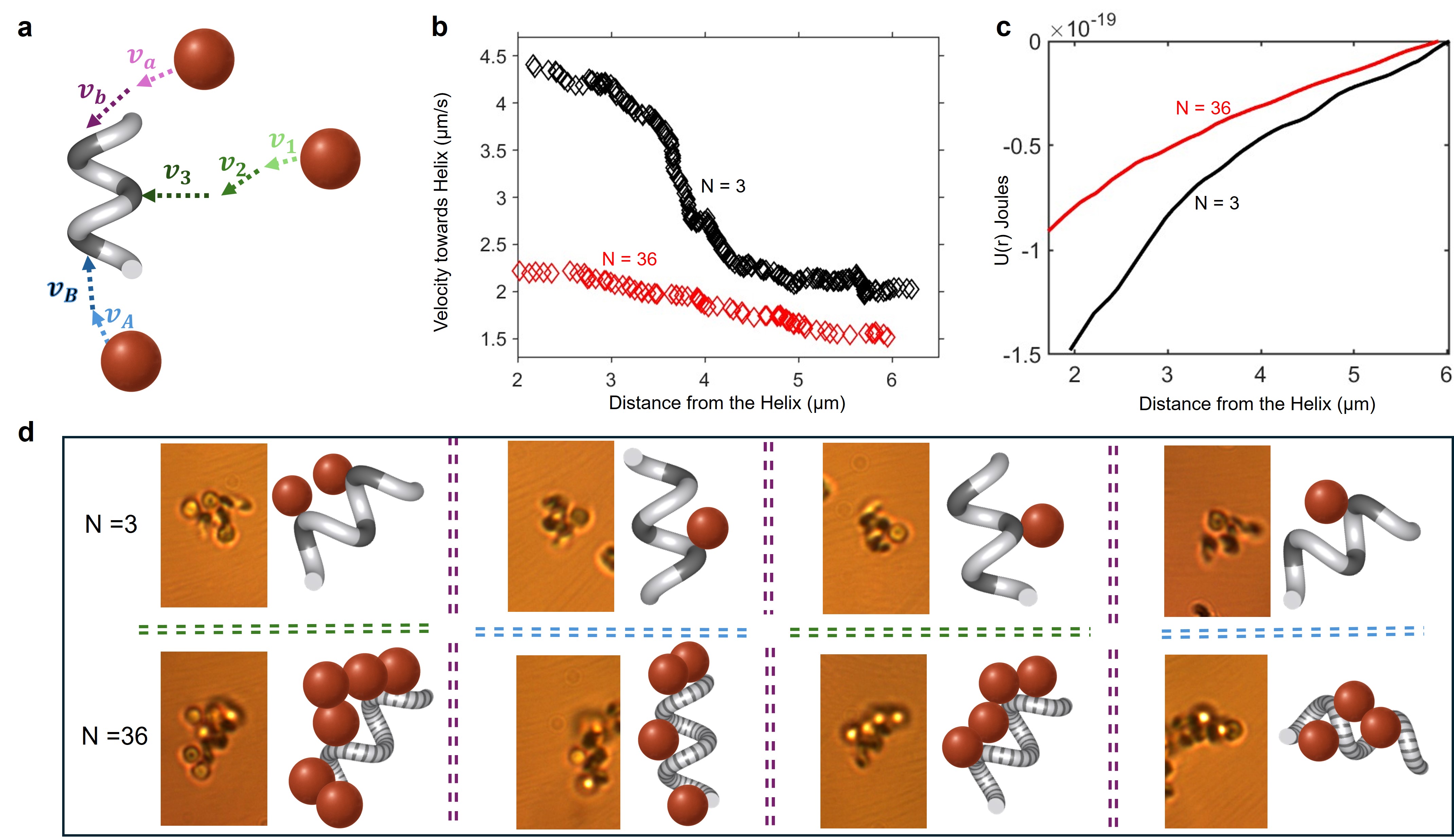}
\caption{\textbf{--- In-situ measurement of the magnetic force landscape
confirms the suppression.} \textbf{a}, Schematic of the experimental
setup: a magnetic microstructure is immobilized on a surface while SPM
beads (probe radius 500~nm) suspended in the fluid migrate toward it
under the magnetic attraction. \textbf{b}, SPM bead velocity toward the
helical microstructure at different separations for $N=3$ and $N=36$.
\textbf{c}, Magnetic interaction energy $U(d)$ between an SPM bead and
the microstructure as a function of separation, reconstructed from the
velocity via Stokes drag. \textbf{d}, For $N=3$ microstructures, SPM
beads localize preferentially at the discrete Fe-rich regions; for
$N=36$, bead attachment spans the entire helical body, indicating a
uniform magnetic force landscape (Movies~S6 and S7).}
\label{fig:5}
\end{figure}

 scanned magnetic cantilever tip\cite{martin1987,ares2015}, establishing that local magnetic fields can be probed directly in fluid. Our measurement
is a colloidal-probe analogue of that approach, a liquid-phase analogue
of atomic force microscopy (Fig.~\ref{fig:5}a). A helical microstructure
is immobilized on the surface of the fluid chamber, SPM beads of radius
$a=500$~nm diffuse in the surrounding fluid, and each bead approaching
the helix drifts under the local magnetic gradient force. Because the
beads move at low Reynolds number, the drift velocity reports the force
directly through Stokes drag, $F(r)=6\pi\eta a\,v(r)$, and integration
yields the pair-interaction energy, $U(r)=\int F\,dr$. This provides a
quantitative, distance-resolved map of the magnetic force and
interaction energy around an individual FM colloid, obtained in the same
fluid environment in which the particle is actuated.
Comparing matched-moment $N=3$ and $N=36$ helices (Fig.~\ref{fig:5}b),
the probe approaching the 3-layer structure accelerates sharply at close
range, whereas near the 36-layer structure its velocity, and hence the
local attraction, remains lower at all measured separations, even though
the total moments are the same. The reconstructed potentials
(Fig.~\ref{fig:5}c) quantify this directly: the 3-layer particle
presents a deep, localized well,
$U_{3}\approx1.4\times10^{-19}~\mathrm{J}$, against the shallower
$U_{36}\approx8\times10^{-20}~\mathrm{J}$ of the 36-layer particle.
Where the beads finally attach is equally telling (Fig.~\ref{fig:5}d):
on $N=3$ structures they localize sharply at the discrete Fe-rich
segments, where the gradient force is largest, whereas on $N=36$
structures attachment is distributed uniformly along the body, the
direct visual signature of a flattened force landscape
(Movies~S6 and S7).
The measured well ratio, $U_{3}/U_{36}$, is a conservative
lower bound on the contact-region suppression for a geometric reason
built into any colloidal probe: the $1/N$ reduction is steepest at
separations comparable to the $\sim7$~nm interlayer spacing, whereas a
500-nm probe cannot approach closer than its own radius and therefore
samples the landscape at a separation where the two architectures have not
yet maximally diverged. The probe measurement thus establishes the
existence, the sign and the spatial structure of the suppression at the
single-particle level.

\subsection*{Application: manipulation in the weak field regime inaccessible
to superparamagnetic colloids}

Stable, high-moment FM colloids allow manipulation in weak field regime; which has been inaccessible to the agglomeration-free colloids in practical use. Weak field typically implies far from the source: a permanent magnet or a gradient coil, produces a field that decays with distance, so the field magnitude $B$ is smallest precisely where the target is deepest, while the gradient $\nabla B$ that
supplies the force decays more slowly. This is where superparamagnetic
colloids, the most widely used agglomeration-free particles, are
intrinsically weak. A particle in a static gradient experiences
$\bm{F}=(\bm{m}\cdot\nabla)\bm{B}$. For an FM colloid of permanent moment
$m$, $F_{\mathrm{FM}}=m|\nabla B|$, set by the gradient alone and
independent of the local field magnitude; for an SPM colloid the moment
is induced, $m_{\mathrm{ind}}=\chi VB/\mzero$, so
$F_{\mathrm{SPM}}=(\chi V/\mzero)\,B\,|\nabla B|$ carries an extra factor
of $B$ that collapses as the field weakens. The drift-speed ratio
therefore scales as $v_{\mathrm{FM}}/v_{\mathrm{SPM}}\propto1/B$, an
inversion: SPM colloids are favoured close to the source, FM colloids
increasingly favoured far from it, exactly where deep, large-volume or
otherwise hard-to-reach targets lie. Because the agglomeration-free
colloids in common use have been SPM or low-moment FM, this is the
region they serve worst, and it has remained out of reach.

\begin{figure}[H]
\centering
\includegraphics[width=\linewidth]{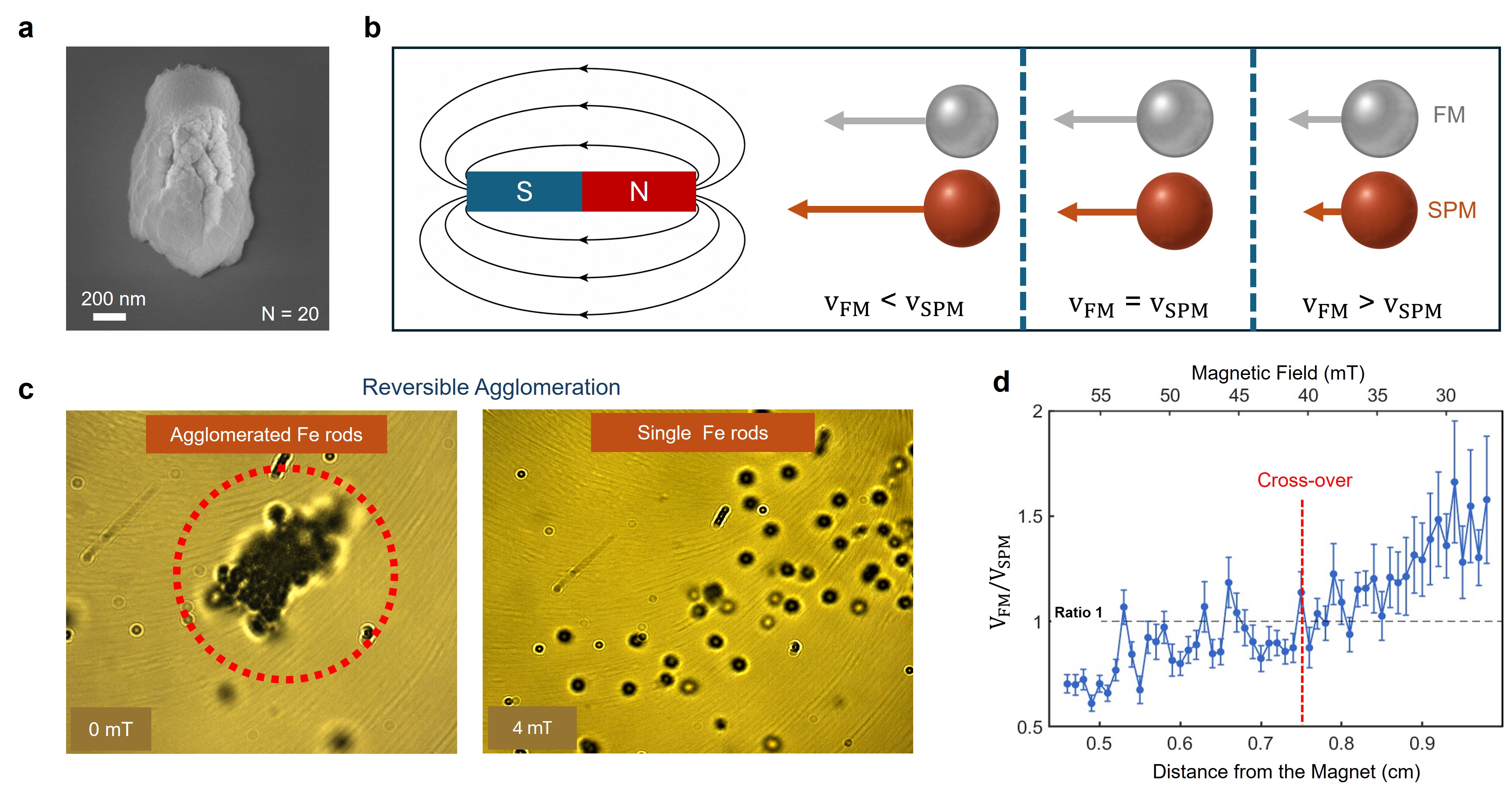}
\caption{\textbf{--- Stable ferromagnetic straight rods reach a
low-field manipulation regime inaccessible to SPM colloids.}
\textbf{a}, SEM of a multilayered Fe/SiO\textsubscript{2} straight rod
($\sim1$~\textmu m length, $\sim600$~nm diameter, $N=20$, layer pitch
$\sim7$~nm). \textbf{b}, Schematic of the gradient-field experiment: a
bar magnet drives both Fe rods (grey) and SPM beads (brown) toward it;
arrow length indicates particle velocity at each distance. \textbf{c},
Representative snapshots showing magnetic agglomeration before field
application and dispersion into individual rods once the field is applied
(Movie~S4). \textbf{d}, Velocity ratio
$v_{\mathrm{FM}}/v_{\mathrm{SPM}}$ as a function of distance from the bar
magnet, demonstrating that the FM rods dominate beyond the crossover
distance of $\approx0.75$~cm.}
\label{fig:6}
\end{figure}

We demonstrate the crossover using multilayered Fe/SiO\textsubscript{2}
straight rods ($\sim1$~\textmu m length, $\sim600$~nm diameter, $N=20$)
benchmarked against commercial 1-\textmu m SPM beads under identical
conditions (Fig.~\ref{fig:6}). 
A NdFeB bar magnet at the wall of a
straight microfluidic channel sets up a gradient field mapped
independently with a Hall probe, and both species are driven toward the
magnet (Fig.~\ref{fig:6}b). The FM rods, agglomerated in zero field,
disperse into individually mobile particles once the field protocol is
applied and remain dispersed throughout (Fig.~\ref{fig:6}c, Movie~S4):
the stability established above is what allows them to act as free
single-particle actuators rather than irreversible clusters. SPM beads
outrun the rods only near the magnet; beyond a crossover distance of
$\approx0.75$~cm the FM rods overtake them and remain faster at all
greater distances (Fig.~\ref{fig:6}d). Because the two species are
matched in size and hydrodynamic drag, the speed ratio in the low-field
region is consistent with the
$v_{\mathrm{FM}}/v_{\mathrm{SPM}}\propto1/B$ scaling. For size-matched
particles the crossover position depends only on the material parameters
$m$ and $\chi$, not on the gradient geometry. Because the crossover distance is already comparable to everyday labware dimensions, the advantage is not confined to specialized deep targets: in a standard cuvette or well most of the volume lies in the weak-field region, where stable FM colloids would outperform the SPM beads currently used for such separations. The benefit of distributed-magnetization colloids therefore extends beyond microrobotics to even routine magnetic-bead separations that are a fixture of everyday laboratory practice.

\section*{Discussion}
\label{sec:discussion}

At the level of design principle, the result is distinct from every
previous route to stabilizing FM colloids. Surface coatings and steric
spacers\cite{spacer} enforce a minimum separation $\delta$ between two
otherwise unmodified dipoles, shifting the pair interaction from
$U\propto\mzero m^{2}/2\pi d^{3}$ to
$\mzero m^{2}/2\pi(d+\delta)^{3}$: the functional form is unchanged, and
the prefactor is reduced only by paying in size. Matching the $N=36$
suppression this way requires $(d+\delta)/d\approx3.3$, roughly tripling
the effective colloid dimension; for rod-like geometries the
spacer-coated equivalent requires an outer diameter $N^{1/3}d$,
inflating the volume by $V'/V\approx N^{2/3}/2$ and raising the
hydrodynamic drag that propulsion and actuation must overcome.
Internal partitioning achieves the same suppression
with the external dimensions essentially preserved, and in the
$N\!\to\!\infty$ limit additionally softens the functional form from
$1/d^{3}$ to a bounded $1/d^{2}$. Superparamagnetic composites, the
other established route, purchase stability by giving up the permanent
moment, and with it precisely the low-field capability demonstrated in
Fig.~\ref{fig:6}.

This work establishes the internal spatial distribution of the
magnetization as a design degree of freedom in its own right, orthogonal
to the material composition, particle shape and surface chemistry that
have so far defined magnetic colloid design. The evidence for its
universality is the phase diagram itself: three particle geometries,
thirteen sample types and two entirely different internal architectures,
discrete layers and dispersed nanoislands, collapse onto a single
boundary governed by $M^{2}/N$. Nothing in the principle is specific to
glancing-angle deposition, which we used for its wafer-scale convenience
and its ability to produce diverse three-dimensional shapes; the same
distributed-magnetization architectures are accessible by
template-directed deposition\cite{physical1,physical2,physical3},
sol--gel co-condensation\cite{refPhilipse,refTartaj} and layer-by-layer
assembly\cite{refDecher}, and we expect the principle to transfer
directly.

By relaxing the de-agglomeration condition $M\Bapp\gtrsim U$, with $U$
falling as $1/N$ at fixed $M$, the magnetization landscape opens an
operating window that has been closed since the constraint was first
articulated: stable, geometrically diverse, high-moment ferromagnetic
colloids, and with them the manipulation regimes that low moments
forbid, low-field gradient targeting far from the source, high-thrust
propulsion at modest applied fields in biological
environments\cite{highthrust1,highthrust2,highthrust3,highthrust4,highthrust5},
and magnetic active matter with strong, permanent, yet switchable
interactions\cite{activematter1,activematter2,activematter3,activematter4,activematter5}. Beyond microrobotics, the reach extends to some of the most common applications as well: even routine magnetic-bead separations can fall in the FM-favoured regime.
More broadly, it reframes the design paradigm for magnetic soft matter: optimizing the internal configuration of the functional material alongside its net composition. We expect the principle to extend naturally to colloids of more complex shapes and field responses.

\section*{Methods}
\label{sec:methods}

\subsection*{Fabrication of helical, bent-rod and straight-rod
microstructures}

Three-dimensional ferromagnetic microstructures of distinct geometries
were fabricated via GLAD\cite{ref13,ref25,refGLAD} onto silicon dioxide
pillar arrays of 400--500~nm diameter. Depositions were carried out in
an electron-beam evaporator at a base pressure of
$\sim10^{-6}$~mbar; deposition rates were $\sim0.5$~\AA~s$^{-1}$ for Fe
and $\sim3$~\AA~s$^{-1}$ for SiO\textsubscript{2}, calibrated
independently with a quartz crystal microbalance. During deposition the
substrate was tilted to an oblique angle of $87\degree$ with respect to
the incident vapour flux, and the rotation speed was precisely
controlled to define pitch and geometry. Helical and bent-rod
architectures were grown to heights of $\approx3$~\textmu m;
straight-rod structures were fabricated to $\approx1$~\textmu m length
and 600~nm diameter, with a diameter distribution of 400--650~nm arising
from broadening effects during deposition. The magnetic component was
incorporated through alternating layer-by-layer Fe/SiO\textsubscript{2}
deposition, with $N$ varied from 1 to 36. In the co-evaporation route,
Fe and SiO\textsubscript{2} were deposited simultaneously at a fixed
Fe:SiO\textsubscript{2} flux ratio.

\subsection*{Per-particle magnetic moment measurement}

Prior to release, all microstructures were magnetized along their short
axis using a calibrated electromagnetic coil, then released into
deionized water by sonication and loaded into sealed glass microfluidic
chambers. Characterization was performed in a custom three-axis
Helmholtz coil assembly mounted on an optical microscope. To extract the
magnetic moment of individual microstructures, the applied field
magnitude was held constant while the driving frequency was swept. The
step-out frequency $\Oc$, beyond which the microstructure loses
synchronization with the rotating field and the propulsion velocity
reaches its maximum, was identified for each swimmer by fitting the
velocity-versus-frequency response. The magnetic moment was estimated
from $\Oc$ using the rotational-drag formalism for anisotropic
geometries, with rotational drag coefficients taken from standard
expressions\cite{ref30}; the moment is related to the step-out
frequency\cite{refGhoshDyn} through $\Oc\propto MB/\gl$, where $\gl$ is
the relevant rotational drag coefficient. This per-particle, in-fluid
protocol avoids the unshadowed Fe growth that causes ensemble
bulk-magnetization measurements (SQUID/VSM) to overestimate the
per-particle moment\cite{ref29}.

\subsection*{In-situ measurement of the magnetic force landscape}

To determine the interaction potential $U(r)$, helical microstructures
with 3 and 36 Fe layers were immobilized on the surface of a
microfluidic glass chamber, and 1-\textmu m SPM beads were dispersed in
water inside the chamber. Beads approaching the helices through Brownian
motion experienced an attractive force arising from the FM
microstructures; their trajectories were tracked and velocities
evaluated as a function of distance. Because the beads move at low
Reynolds number, the net force at any instant is governed by the balance
between attraction and hydrodynamic drag, $F=\gamma v$, where
$\gamma=6\pi\eta a$ is the Stokes drag coefficient. The interaction
potential is obtained by integration, $U(r)=\int F\,dr$. Results for the
3-layer and 36-layer structures provide a direct comparison of
potential-well depths: a deeper minimum corresponds to a stronger
attractive interaction.

\subsection*{Gradient-field manipulation}

Spherical SPM beads of 1~\textmu m diameter and multilayered FM straight
rods ($\approx1$~\textmu m length, 600~nm diameter, $N=20$ alternating
Fe/SiO\textsubscript{2} layers) were suspended in water in a
microfluidic chamber placed on an optical microscope stage. The FM rods
exhibited reversible magnetic agglomeration; clusters dispersed into
individual rods when a rotating field was applied above the
de-agglomeration threshold. A NdFeB permanent bar magnet was positioned
$\approx1$~cm from the field of view, introducing a lateral field
gradient. Both particle types were tracked optically as a function of
distance from the magnet, and the gradient field was calibrated
independently with a Hall-probe sensor. SPM and FM particles were imaged
in separate chambers under identical conditions.

\section*{Acknowledgements}

We thank Sneha G K for the nanopatterned substrates and Ambresh M for
technical support. This work was supported by the Ministry of
Electronics and Information Technology (MeitY), Government of India. We
acknowledge the Centre for Nano Science and Engineering (CeNSE), Indian
Institute of Science, for use of the nanofabrication and
characterization facilities.

\section*{Author contributions}

Paramita Modak carried out the experimental work spanning
nanofabrication, microfluidic characterization, data analysis, and
numerical calculations. Priyadharshini M contributed to part of the
nanofabrication and TEM imaging. Ankur Jaiswar assisted during the
initial phase of the experiments, especially for the low-$N$ multilayer
samples. Ambarish Ghosh supervised the research and contributed to the
overall interpretation of the results and the analytical formulation.
Paramita Modak and Ambarish Ghosh wrote the paper with input from all the authors.

\section*{Competing interests}

The authors declare no competing interests.

\section*{Data availability}

All data supporting the findings of this study are available within the
article. Raw data and analysis code are
available from the corresponding author on reasonable request.

\renewcommand{\refname}{References}

\end{document}